\documentclass[12pt]{article}
\usepackage{graphicx}
\begin{document} 

\title{Market simulation with hierarchical information flux}

\author{Christian Schulze\\
Institute for Theoretical Physics, Cologne University\\D-50923 K\"oln, Euroland}

\maketitle

e-mail: ab127@uni-koeln.de
\bigskip

Abstract: We assume the market price to diffuse in a hierarchical
comb of barriers, the heights of which represent the importance 
of new information entering the market. We find fat tails with
the desired exponent for the price change distribution, and 
effective multifractality for intermediate times.

\bigskip
\bigskip

Among the widely believed stylized facts of real stock markets
are fat tails, corresponding to a cumulative distribution 
function of price changes $r$ decaying as $1/r^3$, and 
multifractality, stating that the exponent of the $q$-th moment 
of this distribution is not a linear function of $q$. We present
here a simple model reproducing both properties.

We start with a hierarchical "comb" of barriers $b(x)$ 
symbolizing information
relevant for the market, as a function of the 
current price $x$ arbitrarily normalized to $0 < x < 1$. If
this unit interval is divided into $2^n$ small intervals 
(we took $n=20$), then at $x = 1/2$ the barrier has height
$b(x)=n$; at $x = 1/4$ and $x = 3/4$ is has $b=n-1$; at 
$x =$ multiples of 1/8 not used before it has $b=n-2$; at
$x =$ multiples of 1/16 not assigned before we set $b = n-3$,
and so on until $b=1$. The remaining $x$ values have $b=0$. 

Now the price $x$, starting in the middle, makes an unbiased
random walk, hindered by the barrier $b(x)$. (More precisely,
$x$ is proportional to the logarithm of the price. Introducing 
a bias did not change out fat tails.)
It overcomes this barrier with a probability exp($-b/2.2)$ and
jumps, in case it overcomes the barrier, by the amount 
$2^{1-n}\exp(b/2.2)$. Thus large $b$ correspond to very rare
and very important informations, shifting the market appreciably,
randomly up or down. (To reduce artificial discontinuities, the
actual barrier heights $b$ where not set to integer values $b=k$
but taken as random between $k-1$ and $k$.)

\vspace*{1cm}
\begin{figure}[htb]
\includegraphics[angle=-90,scale=0.5]{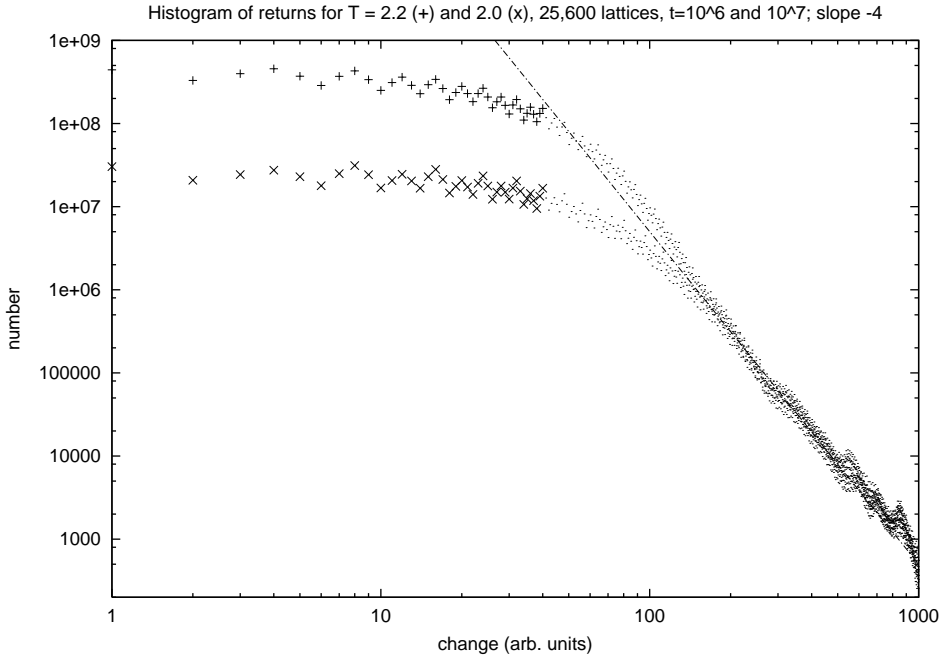}
\caption {Fat tails in the histogram of price changes.}
\end{figure}

Each time step ("day") corresponds to 100 random 
microsteps ("trades") and results in one value $x(t)$ as a 
function of time $t$. The returns $r = x(t+1)-x(t)$ are 
accumulated over $10^6\dots 10^7$ time steps, and about $10^3
\dots 10^4$ independent samples. Figure 1 shows that the 
probability distribution function of these returns follows 
roughly for large $r$ a power law with exponent $-4$, in 
agreement with reality \cite{gopi,lux}. (If we change the
"temperature" parameter 2.2 also this exponent may change.)
By definition, positive and negative changes appear equally often
and equally strong.

For multifractality \cite{muzy}, we generalize the returns to 
$r_\tau = x(t+\tau)-x(t)$ and look at the moments 
$$<r_\tau^q> \, \propto \tau^z $$ 
with exponents $z(q)$ describing the variation of the
returns $r_\tau$ as a function of the time difference $\tau$. 
Fitting these exponents in the time interval $100 < \tau < 1000$,
Fig.2 shows $z(q)$ first to increase with $q$ and then to remain 
roughly constant ("multifractal"), while using less data for 
longer times, $1000 < \tau < 10000$, gives a nearly linear 
increase of $z$ with $q$ ("monofractal"). 

\vspace*{1cm}
\begin{figure}[htb]
\begin{center}
\includegraphics[angle=-90,scale=0.5]{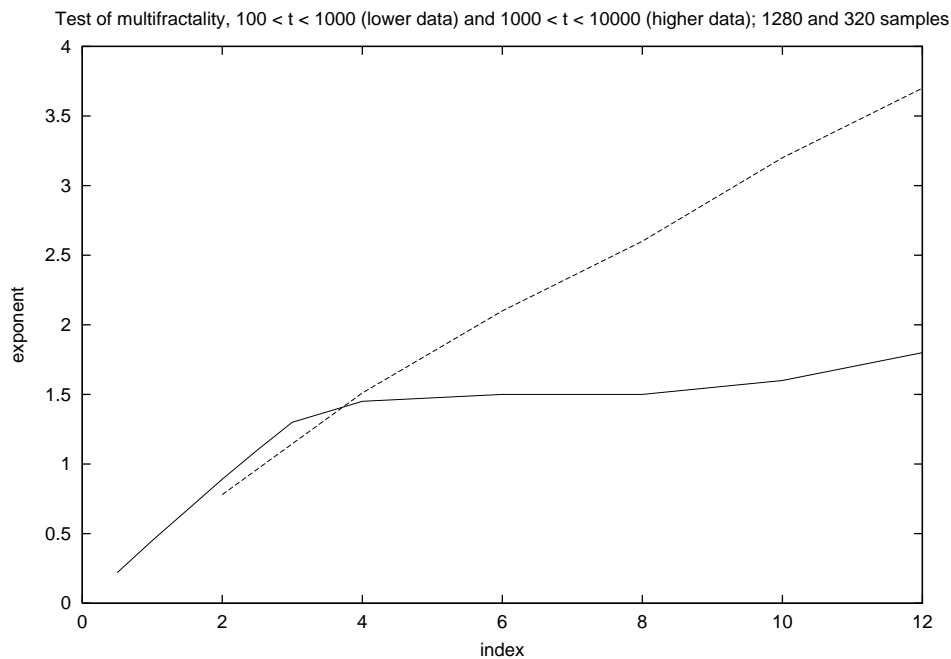}
\caption {Exponent $z$ versus index $q$ for intermediate and
for long times.}
\end{center}
\end{figure}

We thank Deutsche Forschungsgemeinschaft for support, D. 
Stauffer for help with the manuscript, and the supercomputer 
center in Julich for time on the Cray-T3E.

\end{document}